\documentclass[proceedings]{rmaa}

\usepackage{rmaacite}

\renewcommand{\P}[1]{%
\ifnum#1=1\hbox{OW~168--326E}\fi
\ifnum#1=2\hbox{OW~167--317}\fi
\ifnum#1=3\hbox{OW~163--317}\fi
\ifnum#1=5\hbox{OW~158--323}\fi
\ifnum#1=0\hbox{OW~171--334}\fi}

\title{Hipparcos, IUE, and the Stellar Content of the Solar Neighbourhood}
\author{Carlos Allende Prieto
  \affil{McDonald Observatory and Department of Astronomy, The University of Texas at Austin} }

\fulladdresses{
\item Carlos Allende Prieto: McDonald Observatory and Department of Astronomy, The University of Texas at Austin, Austin, TX 78712-1083, USA (callende@astro.as.utexas.edu).}

\shortauthor{Allende Prieto}
\shorttitle{Hipparcos, IUE, and the Solar Neighbourhood}


\abstract{The spectroscopic parallaxes in the Hipparcos catalogue can be used to translate  absolute stellar fluxes observed at Earth to the fluxes emerging at the stellar surface for nearby stars. The comparison of these fluxes with the predictions of theoretical model atmospheres allows us to determine the effective temperature and the metallicity of the stars. It is suggested that is possible to study  the stellar content of the solar neighbourhood making use of the large number of ultraviolet spectra in the archive of the IUE satellite.}

\resumen{La combinaci\'on de flujos absolutos estelares medidos en la Tierra con las pa\-ra\-la\-jes espectrosc\'opicas del cat\'alogo de Hipparcos, permite determinar con precisi\'on los flujos emergentes en las superficies de estrellas cercanas.  Comparando estos flujos con las predicciones de modelos de atm\'osfera te\'oricos, es posible determinar la temperatura efectiva y la metalicidad de las estrellas. Se plantea la posibilidad de estudiar  el contenido estelar de la vecindad solar haciendo uso del gran n\'umero de espectros ultravioletas  disponibles en el archivo del sat\'elite IUE.}  

\listofauthors{C. ~Allende ~Prieto}
\indexauthor{Allende ~Prieto, ~C.}

\begin{document}

\maketitle

\section{Introduction}
\label{sec:intro}
Understanding how the  complex near-UV region of the spectra of stars  is
shaped within the stellar atmospheres can provide a homogeneous source of
information on several of the fundamental stellar parameters, the chemical
composition, magnetic activity, rotational velocity, atmospheric velocities,
and  ages. Many neutron capture elements, whose abundance in
metal-deficient stars keeps log of the supernovae rates along the galactic
evolution, produce undetectable (or subjected to large measurement errors)
features in the optical spectrum, but strong lines in the near-UV (see e.g.
Sneden et al. 1998). Interesting  absorption lines in the spectrum of light atoms, such as boron and beryllium
(see e.g. Garc\'{\i}a L\'opez et al. 1998), are only present in the near-UV. Furthermore, F-type main sequence stars dominate this
region of the spectrum for intermediate-age stellar populations, and
therefore constitute a powerful prospective tool for dating galaxies (Heap
et al. 1999).

A lack of  proper understanding has  prevented full use of the data
gathered by the space observatory IUE to study the stellar population in the
proximity to the Sun. Indeed all the available  determinations of the
metallicity distribution in the  solar neighbourhood, except for the
polemical work by Favata et al. (1997), are based on photometric
calibrations  (Twarog 1980, Wyse \& Gilmore 1995, Rocha-Pinto \& Maciel 1998,
Flynn \& Morell 1997). The star counts do not fit, and the so-called G-dwarf
problem, the deficiency in relative numbers of metal-poor stars in the main
sequence,  seems to extend to other spectral types as well to the more
metal-rich end of the metallicity distribution (Rocha-Pinto \& Maciel 1998).
Ideally, to extract the available information from the near-UV spectrum, the model atmospheres and other spectral synthesis ingredients
should be carefully tuned for each of the stars: the abundances and
abundance anomalies of all relevant elements, microturbulence, effective
temperature, gravity, interstellar extinction, etc. However, as a first step
I am neglecting here the abundance anomalies, variations in microturbulence, and the effect of interstellar extinction, to derive only the
effective temperatures ($T_{\rm eff}$s) and overall metallicities ([Fe/H]s).
Selecting a  sample limited in volume to 100 pc, makes the availability of
Hipparcos (ESA 1997) parallaxes very likely  for the stars observed by IUE,
and places a safe limit on the role of the interstellar extinction.

\section{Observations and analysis}

The Hipparcos Catalogue includes data for 22982 stars within 100 pc
from the Sun. Making use of the MAST Cross Correlation Search Tool, I
have identified 3421 low-resolution LW ($\sim 1800-3500$ \AA) spectra of 992 of
such stars. The IUE (NEWSIPS) observations have been retrieved from the
Villafranca  node of the IUE Final Archive in Spain.  A newer version of the archive has being released
 recently (INES; Rodr\'{\i}guez-Pascual et al. 1999). When more than a single
spectra was available for a given star, they were combined and cleaned
using the IUEDAC IDL Software libraries to produce a single spectrum
per star.

I have made use of the flux distributions calculated by Kurucz, and
available at  CCP7 since 1993. The grid includes models for different
gravities (log g), effective temperatures ($T_{\rm eff}$) and metallicities
([Fe/H]), while the parameters in the mixing-length treatment of the
convection are fixed, as well as it is the microturbulence (2 km/s), and the
abundance ratio between different metals (solar proportions). For a given
set of ($T_{\rm eff}$, log g, [Fe/H]), I obtain the theoretical flux from linear
interpolation, therefore using the information of the eight nearest models
available in the grid, which is divided in steps of 200 K in $T_{\rm eff}$,
0.5 dex in logg, and 0.5 dex in [Fe/H].

Making  use of an accurate stellar parallax (p), $BV$ photometry and
state-of-the-art evolutionary isochrones (Bertelli et al. 1994), one
can estimate the stellar radius (R) with an accuracy of roughly 6\%
(Allende Prieto \& Lambert 2000, Lambert \& Allende Prieto 2000), and
get a small error in the determination of the ratio (pR)$^2$, which
allows to transform the absolute flux measured at Earth to flux
emerging from the stellar atmosphere. It is as well possible to
constrain the mass to within 8\% (in the range of metallicities we are
interested on) and,  therefore,  the gravity within 0.07 dex. Once the
gravity and the dilution factor for the flux are fixed, it is possible to  compare
the  absolute near-UV fluxes  measured by IUE with theoretical fluxes,
and determine $T_{\rm eff}$ and [Fe/H]. As showed by Allende Prieto \&
Lambert (2000), knowing the absolute visual magnitude and the $B-V$ color index, comparison with evolutionary isochrones provides  an independent
estimate of the $T_{\rm eff}$, precise to roughly 2\% for stars with
$4500 < $ $T_{\rm eff}$ $< 8500$ K, which can be used to check for
systematic errors (see below).

\begin{figure}
  \begin{center}
    \leavevmode
    \includegraphics[width=5.5cm,angle=90]{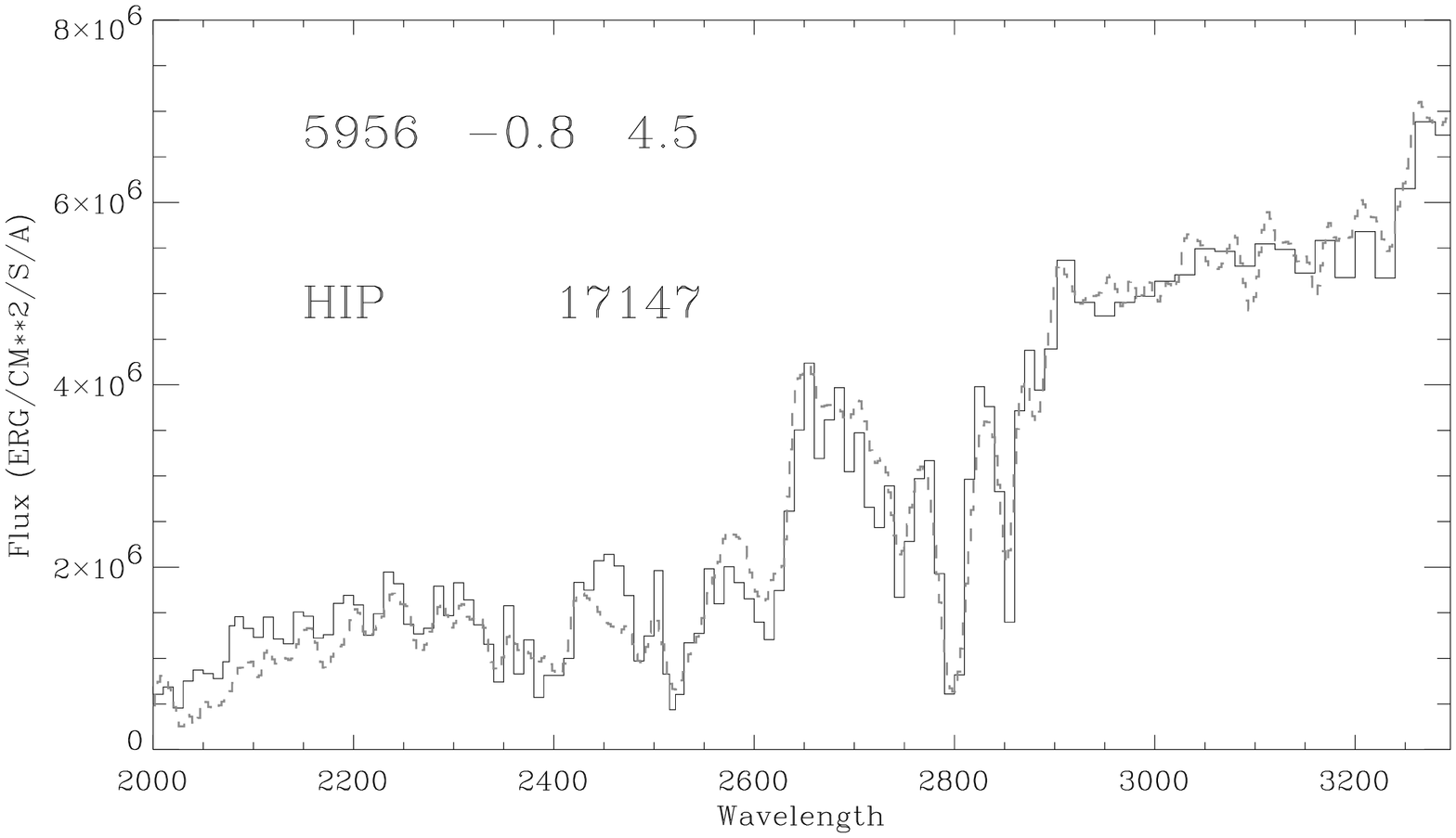}
    \includegraphics[width=5.5cm,angle=90]{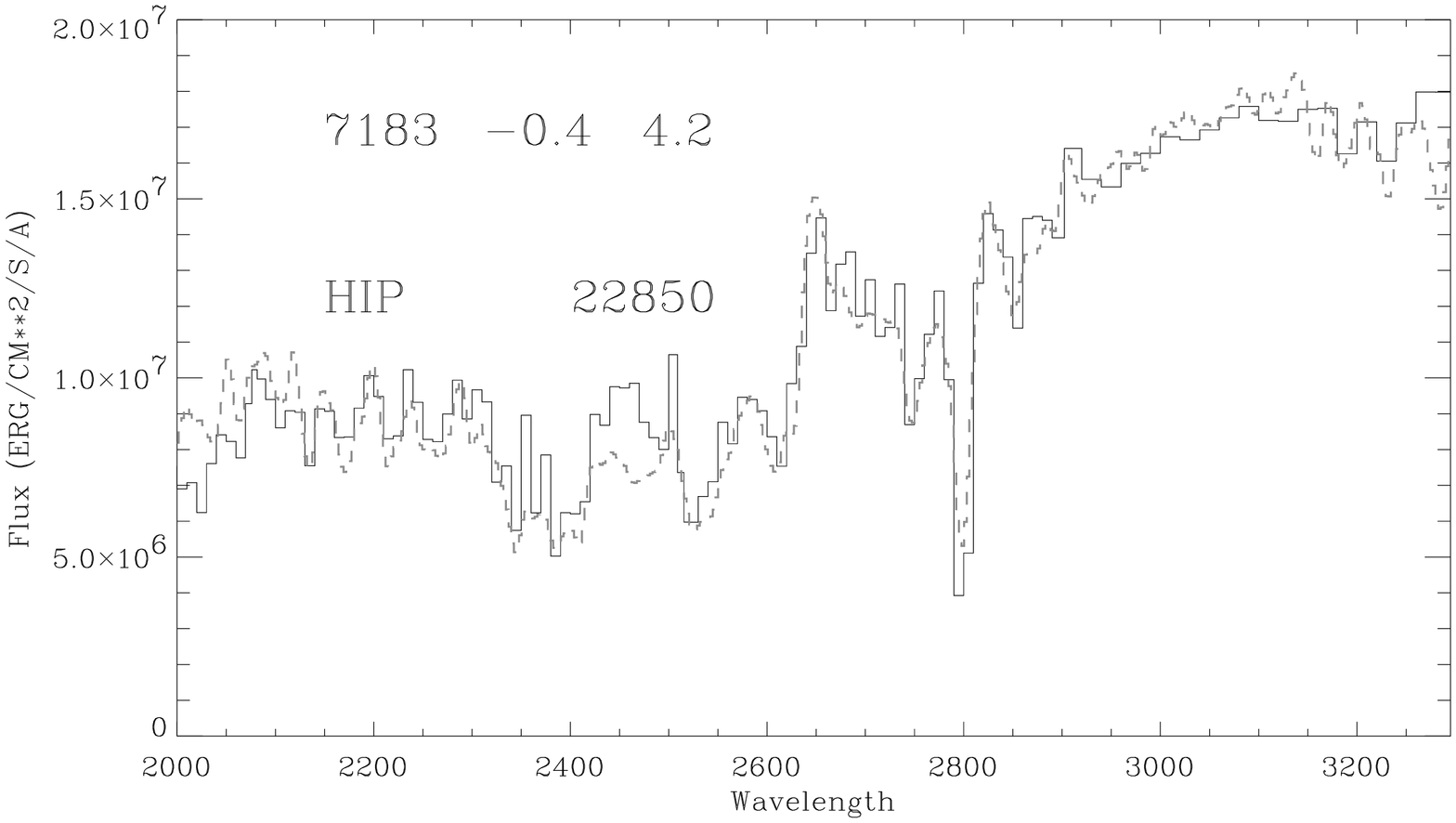}
    \caption{Two examples of the match between observed (dashed lines) and theoretical (solid lines) absolute fluxes at the stellar surface in the region between 2000 and 3000 \AA. Within the panels, $T_{\rm eff}$ (K), [Fe/H] (dex) and logg (g in cgs) have been printed, as well as the star's identification in the Hipparcos catalogue.}
    \label{f1}
  \end{center}
\end{figure}

For each of the analyzed stars I first derive its gravity and the flux
dilution factor, then the values of [Fe/H] and $T_{\rm eff}$  are obtained by finding the minimum of  the square of the difference between the observed and the synthetic spectrum. Previously, the
observed spectrum, which has a resolution between 5.2 and 8.0 \AA\, is degraded
to that of the synthetic spectra, roughly twice poorer. The search for the
optimum  ($T_{\rm eff}$,[Fe/H]) pair is performed using the Nelder-Mead simplex
method, as implemented by Press et al. (1988). Figure \ref{f1} shows two
 examples of the typical goodness-of-fit achieved. 

\section{Discussion}

The sample was restricted to stars with $3500 <$ $T_{\rm eff}$  $<
10000$ K. The employed models are known to be inappropriate close to
and below the cooler limit, as the plentiful molecules are not properly
taken into account,  and the number of nearby stars beyond the upper
limit is very small. 

\begin{figure}
  \begin{center}
    \leavevmode
    \includegraphics[width=5.5cm,angle=90]{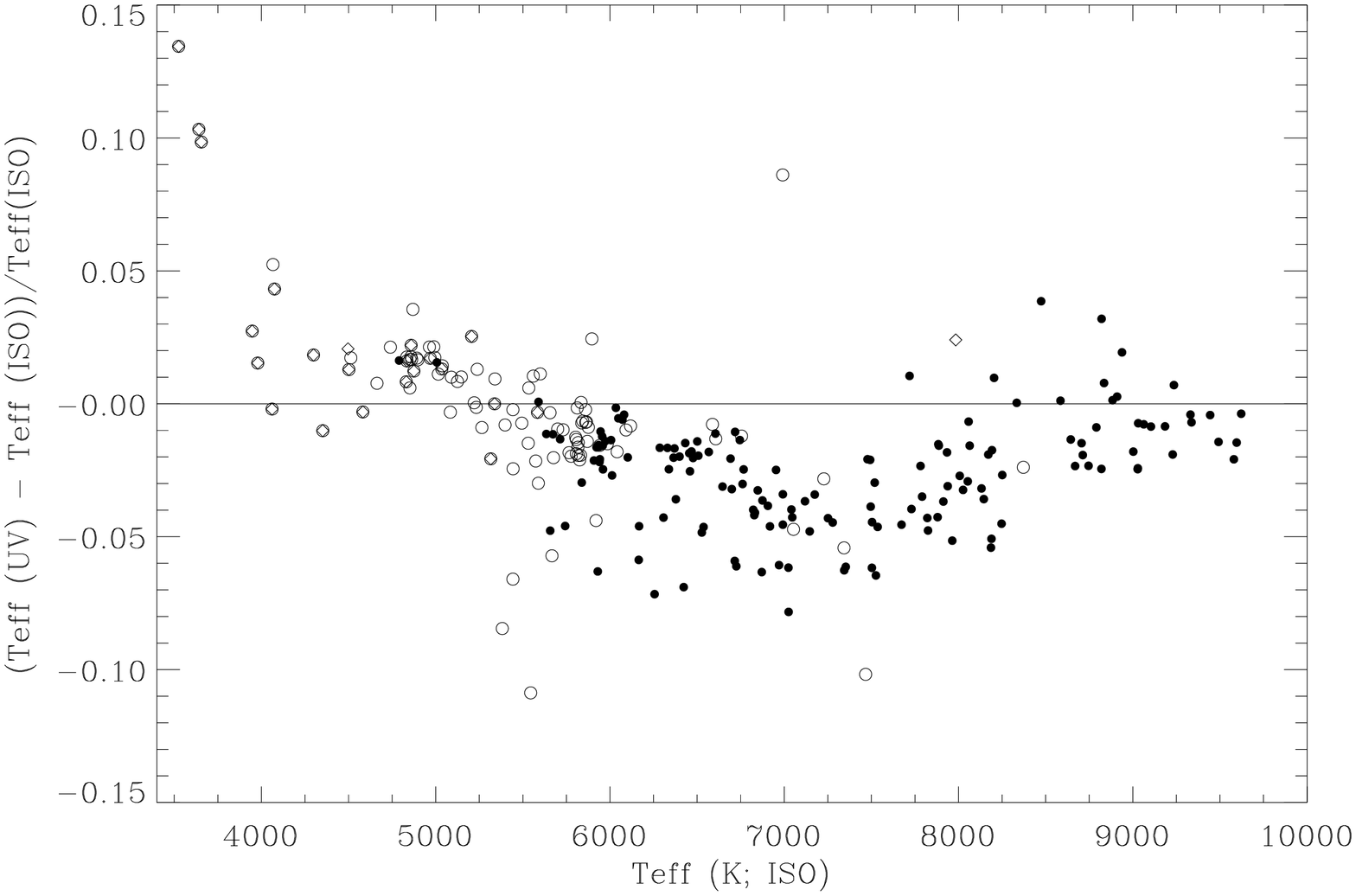}
    \includegraphics[width=5.5cm,angle=90]{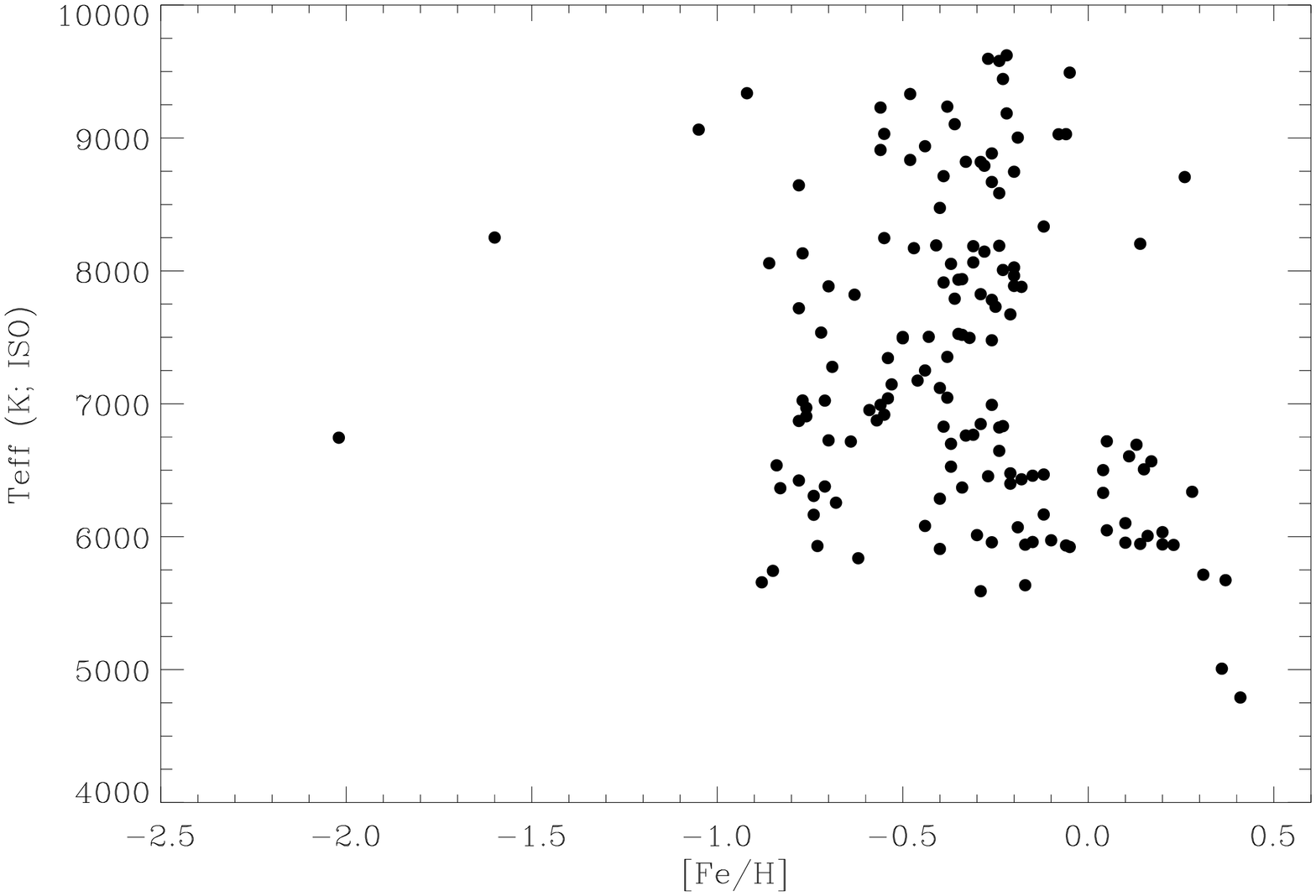}
    \caption{{\it Left panel}: Differences between the effective temperatures derived from the comparison of the stars' position in the colour-magnitude diagram with evolutionary isochrones (labeled ISO) and those from fitting the near-ultraviolet continuum (labeled UV). Stars with strong Mg II 2852 \AA\ emission are identified with rhombi, those with a continuum in the region
2000-2400 \AA\ much stronger than prediction of the models with open circles, and the rest  with filled circles. {Right panel}: Position of the {\it normal} stars (whose spectra do not show evidence for a chromosphere), on the [Fe/H]--$T_{\rm eff}$ plane. Two stars mark off the rest, one is the halo star HD 84937 (HIP 48152; [Fe/H] $\simeq -2.0$) and the other is HD 192640, a well-known $\lambda$ Bootis star. The two hot stars with [Fe/H] $\simeq  -1.0$ are both peculiar too -- HD 161868 is a spectroscopic binary and HD 177724 has been assigned a spectral type A0Vpnn by Abt \& Morrell (1995).}
    \label{f2}
  \end{center}
\end{figure}

A first look at the comparison between the 'evolutionary' and near-UV $T_{\rm eff}$s
reveals systematic differences. The left panel of Figure \ref{f2}  shows the  comparison for the 253 stars whose IUE spectra have perfect quality flags in
the considered spectral range (2000-3000 \AA).  Stars with strong Mg II 2852 \AA\ emission are identified with rhombi, those with a continuum in the region
2000-2400 \AA\ much stronger than the models' prediction with open circles, and
the rest with filled circles. Disregarding the stars for which there is
indication of a chromospheric component in the spectrum (open symbols), the
spectroscopic $T_{\rm eff}$s of 'normal' stars exhibit a systematic difference from
the evolutionary $T_{\rm eff}$s. Similar -- although smaller-- effects could be present
in other $T_{\rm eff}$ scales that make use of the same type of model atmospheres,
such as that derived from the InfraRed Flux Method (IRFM; e.g. Blackwell \&
Lynas-Gray 1994). di Benedeto (1998) showed indeed that the IRFM $T_{\rm eff}$s of A
stars are about 2.3\% lower than his empirical scale. The 150 stars  whose spectra do not show evidence for a
chromosphere are spread out the [Fe/H]--$T_{\rm eff}$ plane as shown in the right panel of Figure \ref{f2}.

It is obviously very difficult to determine the selection effects of the
sample, as  the stars are required to  have been observed by IUE. 
  It is likely that peculiar objects were favoured: multiple systems, stars with abundance
anomalies, pulsating/variable stars, metal-poor stars, etc. That explains
the excess of stars with $-1.0 <$ [Fe/H] $< -0.5$ find here, in comparison with
every other photometric study in the literature. Many of theses stars would
indeed be chemically peculiar stars (not being deficient in all metals, but
 having a low abundance of some of the species relevant to the atmospheric
structure and the near-UV opacities) or spatially-unresolved binary
(multiple) systems. In both cases the stars would tend to be erroneously
 shifted towards hotter $T_{\rm eff}$s and lower  metallicities, in order to
compensate for the excess flux.  The same argument could provide an
explanation for the lack of stars with $T_{\rm eff}$ $> 7000$ K at near/super-solar
metallicities: the peculiar stars, preferred by IUE observers, would tend to
have lower metallicities. Further study is clearly needed to shed light on this issue.

\section{Conclusion}

Several issues have been raised  in this preliminary analysis, and are fundamental. The first of all is
that it is necessary to identify ALL the basic elements (chemical and
non-chemical) that affect the shape of the near-UV spectrum. They may be
quite different for different types of stars. Once those elements are identified,
we should attempt the fit of the spectrum allowing for  ALL to vary. The parameters determined should be compared with other scales, of
empirical or independent nature, to asses the real possibilities of this 
kind of analysis and the performance of the model atmospheres, opacities,
and other modelling ingredients. That procedure will likely help improving
the modelling. It will very likely shed light on another basic problem still
far from solved: the effective temperature scale.

At that stage, we should look back to the application of near-UV
spectroscopic analysis to other affairs, such as the stellar content of
local neighbourhood, and review then the apparent  scarcity of solar and
super-solar metallicity A-F stars.

\acknowledgements 

    I am grateful to Benjam\'{\i}n Montesinos and Enrique Solano  for their help with the IUE spectra, and to David Lambert for interesting discussions.
     This research has made use of the cross-correlation tool of the
     Multimission Archive  at the Space Telescope Science Institute
     (MAST). STScI is operated by the Association of Universities for
     Research in Astronomy, Inc., under NASA contract NAS5-26555.
     Support for MAST for non-HST data is provided by the NASA Office
     of Space Science via grant NAG5-7584 and by other grants and
     contracts. I have made extensive use of  the IDL software libraries 
 developed at the IUE  Data Analysis  Center (IUEDAC). All the data
  used in this work were retrieved from the IUE data server at Villafranca del
 Castillo satellite tracking station of the ESA. The SIMBAD database, operated  at CDS and the  NASA ADS were very useful in this work.



\begin{thebibliography}

\bibitem[<>]{} {Abt}, H. ~A.  \& {Morrel}, N. ~I. 1995, ApJS  {99}, 135

\bibitem[<>]{}
 {Allende Prieto}, C. \& {Lambert}, D. ~L. 1998, A\&A {352}, 555

\bibitem[<>]{}
 {Bertelli}, G., {Bressan}, A., {Chiosi}, C., {Fagotto}, F. \& {Nasi}, E. 1994, A\&AS {106}, 275

\bibitem[<>]{}
 {Blackwell}, D. ~E. \& {Lynas-Gray}, A. ~E. 1994, A\&A {282}, 3


\bibitem[<>]{}
 {di Benedetto}, G. ~P. 1998, A\&A {339}, 848

\bibitem[<>]{}
 {ESA} 1997, The Hipparcos and Tycho Catalogues (ESA SP-1200)

\bibitem[<>]{}
 {Favata}, F., {Micela}, G. \& {Sciortino}, S. 1997, A\&A  {323}, 809

\bibitem[<>]{}
 {Flynn}, C. \& {Morel}, O. 1007, MNRAS {286}, 617

\bibitem[<>]{}
 {Garc\'{\i}a L\'opez}, R. ~J., {Lambert}, D. ~L., {Edvardsson}, B., {Gustafsson}, B., {Kiselman}, D. \& {Rebolo}, R. 1998, ApJ  {500}, 241

\bibitem[<>]{}
 {Heap}, S., {Brown}, T., {Lanz}, T. \& {Yi}, S. 1999,  Ap\&SS {265},  531 


\bibitem[<>]{}
 {Lambert}, D. ~L. \& {Allende Prieto}, C. 2000, In: R.~J. Garc\'{\i}a L\'opez, R. Rebolo \&  M.~R. Zapatero Osorio (eds.): {The 11$^{\rm th}$ Cambridge Workshop on Cool Stars, Stellar
Systems and the Sun},  ASP (Conf. Series),  in press (Available at
  http://hebe.as.utexas.edu/cs11rt/rt.html)



\bibitem[<>]{}
 {Press}, W.~H., {Flannery}, B.~P., {Teukolsky}, S.~A. \& {Vetterling}, W.~T. 1988, {Numerical Recipes} (Cambridge: Cambridge Univ. Press)

\bibitem[<>]{}
 {Rocha-Pinto}, H.~J. \& {Maciel}, W.~J. 1998, A\&A {339}, 791

\bibitem[<>]{}
 {Rodr\'{\i}guez-Pascual}, P.~M., {Gonz\'alez-Riestra}, R., {Schartel}, N. \& {Wamsteker}, W. 1999, A\&AS {139}, 183

\bibitem[<>]{} 
 {Sneden}, C., {Cowan}, J.~ J., {Burris}, D. ~L. \& {Truran}, J. ~W. 1998, ApJ {496}, 235

\bibitem[<>]{}
 {Twarog}, B. A. 1980, ApJ {242}, 242


\bibitem[<>]{}
 {Wyse}, R. ~F.~ G. \& {Gilmore}, G. 1995, AJ {110}, 2771

\end{thebibliography}
\end{document}